\documentclass[utf8]{frontiersSCNS} 

\usepackage{url,lineno,microtype,subcaption}
\usepackage[onehalfspacing]{setspace}
\usepackage{graphicx}

\newcommand{\rev}{\textcolor{black}}
\usepackage{color,amssymb}
\usepackage{float}
\usepackage{amsmath}
\usepackage{amssymb}
\usepackage[french,english]{babel}

\def\keyFont{\fontsize{8}{11}\helveticabold }
\def\firstAuthorLast{Scheibert {et~al.}} 
\def\Authors{Julien Scheibert\,$^{1,*}$, Riad Sahli\,$^{1}$ and Michel Peyrard\,$^{2}$}

\def\Address{$^{1}$Univ Lyon, Ecole Centrale de Lyon, ENISE, ENTPE, CNRS, Laboratoire de Tribologie et Dynamique des Systèmes
LTDS UMR5513, F-69134, Ecully, France \\
$^{2}$Universit\'e de Lyon, Ecole Normale Sup\'erieure de Lyon, Laboratoire de Physique, CNRS, UMR 5672, 46 all\'ee d’Italie, F-69364 Lyon Cedex 7, France}

\def\corrAuthor{J. Scheibert}

\def\corrEmail{julien.scheibert@ec-lyon.fr}

\begin{document}
\onecolumn
\firstpage{1}

\title[Onset of sliding of elastomer multicontacts: an asperity model]{Onset of sliding of elastomer multicontacts: failure of a model of independent asperities to match experiments} 

\author[\firstAuthorLast ]{\Authors} 
\address{} 
\correspondance{} 

\extraAuth{}

\maketitle

\begin{abstract}

Modelling of rough frictional interfaces is often based on asperity models, in which the individual behaviour of individual microjunctions is assumed. In the absence of local measurements at the microjunction scale, quantitative comparison of such models with experiments is usually based only on macroscopic quantities, like the total tangential load resisted by the interface. Recently however, a new experimental dataset was presented on the onset of sliding of rough elastomeric interfaces, which includes local measurements of the contact area of the individual microjunctions. Here, we use this more comprehensive dataset to test the possibility of quantitatively matching the measurements with a model of independent asperities, enriched with experimental information about the area of microjunctions and its evolution under shear. We show that, despite using parameter values and behaviour laws constrained and inspired by experiments, our model does not quantitatively match the macroscopic measurements. We discuss the possible origins of this failure.

\tiny
 \keyFont{ \section{Keywords:} rough contact, elastomer friction, onset of sliding, asperity model, shear-induced area reduction, stick-slip, elastic interactions} 
\end{abstract}

\section{Introduction}
\label{sec:intro}
The mechanical behaviour of contact interfaces between rough solids is crucial to understand their tribological properties. The rough contact mechanics community has been developing models in two main directions (see~\cite{vakis_modeling_2018} for a recent review). First, \textit{asperity models} in which the contact interface is divided into well-defined microjunctions actually carrying the normal and tangential loads applied to the contacting solids~\citep{braun_transition_2002,ciavarella_re-vitalized_2006,violano_contact_2019}. Each microjunction is ascribed a set of individual properties (\textit{e.g.} its height, radius of curvature or friction coefficient) necessary to apply some assumed behaviour laws (\textit{e.g.} any contact~\citep{johnson_contact_1987} or friction law~\citep{le_bot_relaxation_2019}) when submitted to an external stimulus. The macroscopic behaviour of the interface is then the emerging, collective response of the population of microjunctions~\citep{tromborg_slow_2014,braun_seismic_2018,costagliola_2-d_2018}. Second, \textit{continuum models} in which the input quantity is the full topography of the rough surfaces, and an exact solution of the unilateral contact and friction problem is seeked~\citep{pastewka_contact_2014,yastrebov_role_2017,ponthus_statistics_2019}, again under some assumptions on the interfacial behaviour, concerning \textit{e.g.} elasticity, friction and adhesion.

Each approach can be used to produce two types of results, either deterministic or statistical. Deterministic results are obtained for a given topography (for continuum models) or a given set of model parameters (for asperity models, including the properties of each microjunction) and are thus specific to those input data. They are relevant for quantitative comparison with a particular experiment. In contrast, statistical results are the expected results of a large number of deterministic calculations performed on statistically similar random surfaces. In asperity models, statistical results are obtained when using probability density functions (pdfs) of the microjunction properties~\citep{greenwood_contact_1966,braun_modeling_2008,thogersen_history-dependent_2014}. In continuum models, they are usually obtained using the power spectrum density (psd) of the topographies under study~\citep{persson_theory_2001}. In the following, we aim at finding a quantitative match with a specific set of measurements, so we will consider deterministic results.

\begin{figure}[tb!]
\includegraphics[width=\columnwidth]{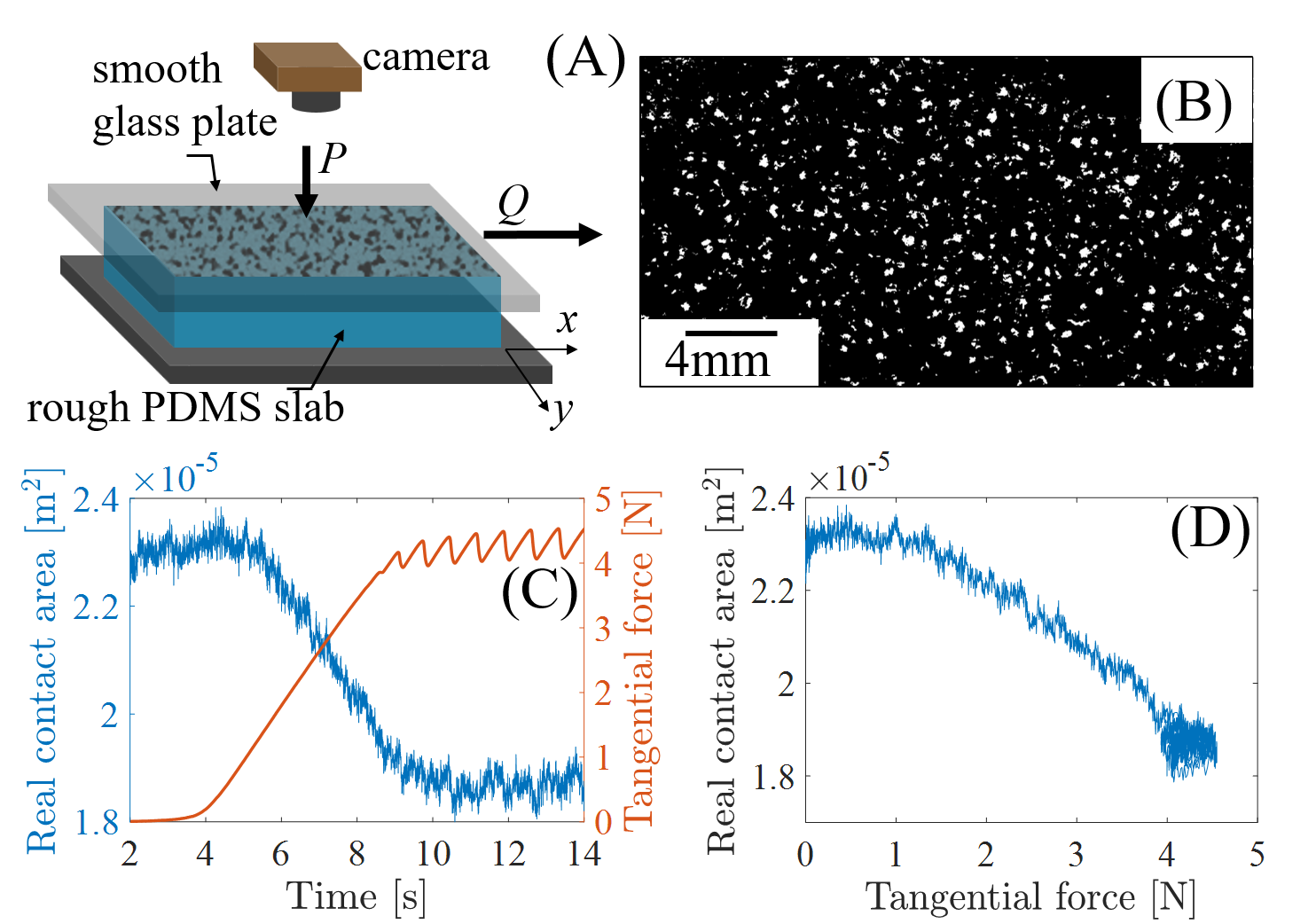}
\caption{Experiments that our model attempts to reproduce. A: Sketch of the experimental setup. B: Typical segmented image of the interface showing individual microcontacts in white. $P$=6.40N. C: Concurrent time evolutions of the tangential force $Q$ (red) and the area of real contact (blue), for $P$=3.10N. D: Area of real contact as a function of the tangential force, for the same data as in C.}
\label{Fig:expmacro}
\end{figure}

Both asperity and continuum models have been widely explored in the context of rough contacts under purely normal load, with a recent study explicitly comparing the relative merits of the two approaches~\citep{muser_meeting_2017}. Several studies aimed at a quantitative comparison between deterministic model results and local, microjunction level measurements (see \textit{e.g.} ~\cite{acito_adhesive_2019,mcghee_contact_2017}). In contrast, to our best knowledge, such comparisons have not been reported in the case of sheared multicontacts. Here we will attempt to build an asperity model able to quantitatively match recent measurements performed on the incipient tangential loading and onset of sliding of a rough elastomer slab in contact with a smooth glass plate~\citep{sahli_evolution_2018,sahli_shear-induced_2019} (Fig.~\ref{Fig:expmacro}A). Those measurements (see a typical example in Fig.~\ref{Fig:expmacro}C,D) are particularly interesting and constraining for models because, in addition to the macroscopic loads on the interface, they include the evolution under shear of the individual contact areas and shapes of the many microjunctions forming the interface (Fig.~\ref{Fig:expmacro}B).

The philosophy of this work is to start with a model of independent asperities like the earthquake model of~\cite{braun_modeling_2008}, enrich it with the recently identified phenomenology of shear-induced area reduction, and genuinely ask the question whether such a model is sufficient to quantitatively match a particular experimental dataset. In other words, we do not aim at a definitive model of the incipient tangential loading and onset of sliding of rough elastomer contacts. Rather, we make one single step ahead compared to the models in the literature, and try to conclude whether this step (including shear-induced area reduction) is sufficient or not. Such an approach can only be fruitful if the values of the model parameters are sufficiently constrained by the experimental dataset, so that one avoids fortuitous agreement. This can be achieved (i) by limiting to the strict minimum the number of parameters that cannot be directly measured experimentally, and (ii) by performing a thorough exploration of the parameter space for those remaining unconstrained parameters. In this work, we did our best to apply this strategy, which in our case leads to an unsatisfactory agreement. This result is nevertheless a progress in the sense that it clarifies the range of assumptions that remain to be questioned and improved in future studies.

In section~\ref{sec:model}, we describe the asperity model and provide the experimental constraints on the model parameters in section~\ref{sec:expparam}. Quantitative comparisons between the model and measurements are given in section~\ref{sec:compa}, while in section~\ref{sec:discussion} we discuss the possible reasons for the absence of good matching between the two.

\section{Materials and methods}
\subsection{Model description}\label{sec:model}

We consider the frictional interface between a slider of mass $M$ and a track. The tangential displacement of the slider, $X(t)$, is assumed to obey the following equation of motion: 
\begin{equation}
  \label{eq:motion}
  M \ddot X + M \eta \dot X =  k_L (v t - X) - F,
\end{equation}
where $M$ is the slider's mass, $k_{L}$ is the stiffness of the loading spring through which the slider is pulled at constant velocity $v$, $\eta$ is an \rev{effective} viscous parameter \rev{accounting for dissipation \textit{e.g.} in the air or in the loading spring}, the dot indicates the time derivative, and $F$ is the resistive force due to interfacial\rev{/adhesive} friction.

We assume that the interface is a multicontact made of $N$ independent individual microjunctions, each resisting a force $f_i$, so that $F=\sum_{i=1}^{N} f_i$. Each microjunction can be in either of two states. First, a \textit{pinned} state during which the junction acts like a (\rev{time-evolving}) elastic spring of stiffness $k_i$, so that $f_i=k_i (X(t)-x_i)$, with $x_i$ the slip displacement of the junction with respect to the track (\textit{e.g.} $x_i = 0$ as long as junction $i$ has never been slipping). When a threshold force $f_{si}$ is reached, the junction enters a \textit{slipping} state, during which $f_i = \epsilon \, f_{si}$. Note that $\epsilon < 1$, so that $f_{si}$ and $\epsilon \, f_{si}$ are the analogs, at the junction level, of a static and a dynamic friction force, respectively.

The mechanical behaviour of individual junctions is inspired by experimental observations made on the same setup and materials in contact as in Fig.~\ref{Fig:expmacro}A, but when the rough slab is replaced by a single smooth sphere~\citep{sahli_evolution_2018,sahli_shear-induced_2019,mergel_continuum_2019}. The resulting sphere/plane contact is assumed to be representative of an individual microjunction within a multicontact like that of Fig.~\ref{Fig:expmacro}B. Those experiments, carried out both for large normal loads~\citep{sahli_shear-induced_2019} and for small (even negative) normal loads~\citep{mergel_continuum_2019}, have shown that, under increasing shear, the initially circular contact shrinks anisotropically and becomes increasingly ellipse-like. As shown in~\cite{sahli_shear-induced_2019} the shrinking minor axis of the ellipse is parallel to the shear loading direction, while the variations of the major axis (in the direction orthogonal to shear) can be neglected.

Defining $\ell_{\parallel i}$ and $\ell_{\perp i}$ the size of an elliptic microjunction along and orthogonal to shear, respectively, we can define its area as $A_i=\frac{\pi}{4} \ell_{\parallel i} \ell_{\perp i}$. Following~\cite{mindlin_compliance_1949}, the stiffness of \rev{such an elliptic} contact along the shear direction is, \rev{assuming no-slip contact conditions:}
\begin{equation}
\label{eq:stiffness}
k_i= \frac{ \frac{\pi}{2} \ell_{\perp i} E}{(1 + \nu)\left[ \bold{K}(e) - \frac{\nu}{e^2} \Big( \bold{K}(e) - \bold{E}(e) \Big) \right] },
\end{equation}
with $E$ and $\nu$ being the Young's modulus and Poisson's ratio of the material that constitutes the microjunctions, $e=\sqrt{1 - \frac{\ell_{\parallel i}^2}{\ell_{\perp i}^2}}$ is the excentricity of the junction, $\bold{K}$ and $\bold{E}$ are the elliptic integrals of the first and second type, respectively. \rev{Note that assuming that micro-junctions are elliptic is the simplest increment of realism compared to a circular assumption, in order to account for the complex shapes observed for micro-junctions in the experiments.}

Assuming that each microjunction is initially circular, we can define the common initial value, $\ell_{0 i}$, of $\ell_{\perp i}$ and $\ell_{\parallel i}$ from \rev{its initial individual} area $A_{0 i}$ as $\ell_{0 i}=\sqrt{\frac{4 A_{0i}}{\pi}}$. As already mentioned, $\ell_{\perp i}$ varies negligibly under shear, so we will consider that $\ell_{\perp i}=\ell_{0 i}$ at all times. The evolution of \rev{each} $\ell_{\parallel i}$ is then deduced from the shear-induced area reduction reported in~\cite{sahli_evolution_2018}:
\begin{equation}
  \label{eq:areadecay}
  A_i = A_{0i} - \alpha_b \frac{1}{A_{0i}^p} \; f_i^2,
\end{equation}
with $\alpha_b$ and the exponent $p$ two constant parameters of the model. The \rev{size of junction $i$} along the shear direction is thus simply $\ell_{\parallel i}=\frac{4 A_i}{\pi \ell_{0i}}$. \rev{Note that there is currently no rigourous contact mechanics theory for the evolution of the shear stiffness of a sheared sphere/plane contact that would incorporate anisotropic contact area reduction. Here, such a behaviour is approximated at all times by the combination of Eq.~\ref{eq:stiffness}, which is valid under no-slip assumption, and of Eq.~\ref{eq:areadecay}, which was empirically found at macroscale. Doing so, we assume that Eq.~\ref{eq:areadecay} also applies at microscale, as suggested by the existence of common values of $\alpha_b$ and $p$ for both the macro- and micro-scales~\citep{sahli_evolution_2018}.}
 
\begin{figure}[tb!]
\begin{centering}
\includegraphics[width=0.7\columnwidth]{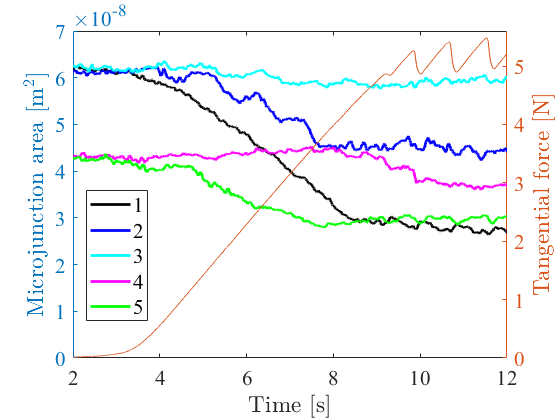}
\caption{Left axis: time evolution of the contact area $A_i$ of 5 typical microjunctions in the experiment at $P$=4.01 N. Right axis: concurrent evolution of the tangential force $Q$.}
\label{Fig:microjunctions}
\end{centering}
\end{figure}

For each microjunction, Eq.~(\ref{eq:areadecay}) is used from the beginning of the experiment, when $f_i$ assumed to be 0, up to when the junction first starts to slip (when $f_i=f_{si}$). At that instant, $A_i$ takes the value $A_{si}=A_{0i} - \alpha_b \frac{1}{A_{0i}^p} \; f_{si}^2$. For later times, based on the observation of the typical behaviour of $A_i$ during the experiments of~\cite{sahli_evolution_2018} (see Fig.~\ref{Fig:microjunctions}), we assume that $A_i$ always remains equal to $A_{si}$.

In contrast, the force resisted by a microjunction can vary with time after the first onset of slipping. When the slider's velocity, $\dot X(t)$, gets smaller than a minimum value $\dot X_{\mathrm{min}} = c_{\mathrm{min}} \times v$, with $c_{\mathrm{min}}$ a scalar parameter, we assume that all the slipping contacts will repin, with a position $x_i = X - \epsilon f_{si} / k_i $. Doing so, at repinning, there is no force discontinuity, as the repinning force $k_i (X-x_i)$ is equal to the slipping force $\epsilon f_{si}$.

Following~\cite{sahli_evolution_2018}, the threshold force $f_{si}$ at which the microjunction starts to slip is assumed proportional to its area at the same instant, \textit{i.e.} $f_{si}=\sigma A_{si}$, with $\sigma$ the frictional shear strength of the contact.

The algorithm used to solve the model numerically is provided in Appendix 1.

\subsection{Experimental constraints}\label{sec:expparam}
In order to quantitatively compare the model results with the multicontact experiments reported in~\cite{sahli_evolution_2018,sahli_shear-induced_2019}, we need to feed the model with parameter values based on the measurements. In the following list, we first provide all constant parameter values that are directly accessible experimentally, with the error bars when relevant.
\begin{itemize}
\item $M=M_{0}+M_{1}$, where $M_{0}$=100 g is the mass of the slider and $M_{1}$=0, 55, 111, 215, 308, 552 g an additional mass, for the 6 experiments performed. All masses are given at $\pm$1g.
\item $k_L$=9200$\pm$200 N$/$m~\citep{sahli_evolution_2018}.
\item $v$=0.1 mm$/$s.
\item $E$=1.6$\pm$0.1 MPa~\citep{sahli_evolution_2018}. This value and the associated error bar are the mean value and standard deviation over 32 estimates using 5 different spherical PDMS samples prepared in the same conditions.
\item $\nu$ is assumed to be equal to 0.5, as is classically done for elastomers.
\item the \rev{individual values of the initial areas of all micro-junctions,} $A_{0i}$, are extracted from the initial image (for $Q$=0), segmented as described in~\cite{sahli_evolution_2018}. \rev{The fact that all micro-junctions have different areas is the result of the random nature of the elastomer surface topography (see a typical Power Spectrum Density in Supplemental Material of Ref.~\citep{sahli_shear-induced_2019}) and of its elastic contact interaction with the rigid glass plate.}
\item $N$ is also extracted form the same segmented image.
\item $\sigma_{exp}$=0.23$\pm$0.02 MPa~\citep{sahli_evolution_2018}, is the experimental value of the frictional shear strength of the interface, determined from a linear fit of $(A_s,Q_s)$ for the 6 experiments. $Q_s$ is the macroscopic static friction (peak) force and $A_s$ is the total area of real contact at the same instant. We will discuss below how the value of $\sigma$ in the model is related to $\sigma_{exp}$.
\end{itemize}

There are three model parameters which cannot be directly measured in experiments: $\eta$, $\epsilon$ and $c_{min}$.

$\eta$ is introduced to enable energy dissipation in the system, thus avoiding spurious oscillations of the slider. However, the value of $\eta$ should not be too large, because it would prevent the possibility of stick-slip in the model. We found that stick-slip exists up to $\eta$ between 180 and 200, but for those large values, the initial stick-slip cycles are significantly different from the experiments. In practice, we found that
\begin{equation}
\eta=100
\end{equation}
is a good compromise between oscillation reduction and a reasonable reproduction of the stick-slip sequence. The results are rather insensitive to the precise value of $\eta$, since $\eta$=50 gives virtually identical results.

$\epsilon$ has a leading order control on the amplitude and period of the tangential force fluctuations during stick-slip. Systematic tests of the model for various values of $\epsilon$ led us to choose
\begin{equation}
\epsilon=0.90.
\end{equation}
In particular, this value is sufficiently small to enable stick-slip for all 6 normal loads (as observed experimentally), while reproducing reasonably well the amplitude and period of the stick-slip sequences in all cases.

Note that in the model, if there was no stick-slip, the steady-state sliding friction force would be $\sum_{i=1}^N \epsilon \sigma A_{si}$ (all microjunctions are in their slipping state). In order for this value to match the macroscopically measured value $Q_s=\sigma_{exp} A_s$, one has to impose that
\begin{equation}
\sigma=\frac{\sigma_{exp}}{\epsilon},
\end{equation}
and this is what we do in the following.

Our tests showed that the value of $c_{min}$ has no impact on the results as far as it is sufficiently small. For instance, simulations with $c_{min}$=10$^{-5}$ are essentially undistinguishable from those using 0.01. The reason is that, when $\left|\dot{X}(t)\right|$ crosses the value $c_{min} \times v$, the velocity drop is so fast that the time at which the crossing occurs is almost independent on the value of $c_{min}$. In our calculations, we will use
\begin{equation}
c_{min}=0.01.
\end{equation}

Extracting values for $p$ and $\alpha_b$ in Eq.~(\ref{eq:areadecay}) requires fitting the power law relationship between the individual area reduction parameters, $\alpha_i$, and the initial areas, $A_{0i}$, presented as purple squares in Fig.~3 of~\cite{sahli_evolution_2018}. Such a fitting is actually difficult due to the large dispersion of the data, as can be inferred from the large difference in total area decay of microjunctions 1 to 3 in Fig.~\ref{Fig:microjunctions}, although they start with almost identical areas. A fit letting both $\alpha_b$ and $p$ as fitting parameters gives 95$\%$ confidence error bars as large as 600$\%$ for the optimum value for $\alpha_b$, which is not a viable option. We then tried to fix the value of $p$ and fit the data with $\alpha_b$ being the only fitting parameter. We found that the quality of the fit (quantified by its $R^2$ value) was essentially independent of $p$ (as long as it is not too different from the value 3/2 proposed in~\cite{sahli_evolution_2018}), preventing any objective choice of $p$.

Based on those observations regarding the determination of $p$ and $\alpha_b$ from experimental data, in our model studies we decided to fix $p$ and, for each value of $p$, we determined the value of $\alpha_b$ that gives the best agreement between the area decay predicted by the model and that measured in the experiments. To do that, we fitted both the experimental and model version of the curve $A(Q)$ by a quadratic function of the form $A=A_0-\alpha Q^2$. $A_0$ being the same in the model as in the experiment (because $A_0=\sum_{i=1}^N A_{0i}$), the fitting procedure enables identification of an $\alpha_b$ which provides an exact match between the two quadratic decays. Importantly, we found that, for all tested values of $p$ close to 3/2 (the value suggested in~\cite{sahli_evolution_2018}), the model results (when using the corresponding fitted $\alpha_b$) were almost undistinguishable. So, in practice, we chose $p$=3/2, for which the model studies give an optimal $\alpha_b$=0.45 10$^{-15}$ m$^{5}$/N$^{2}$ for the experiment with the smallest normal load, and $\alpha_b$=1.00 10$^{-15}$ m$^{5}$/N$^{2}$ for the experiment with the largest normal load. We then adopted the average value between both, $\alpha_b$=0.725 10$^{-15}$ m$^{5}$/N$^{2}$, as a constant to be used for all experiments.

\section{Results: quantitative comparison}\label{sec:compa}
\begin{figure}[tb!]
\begin{centering}
\includegraphics[width=0.7\columnwidth]{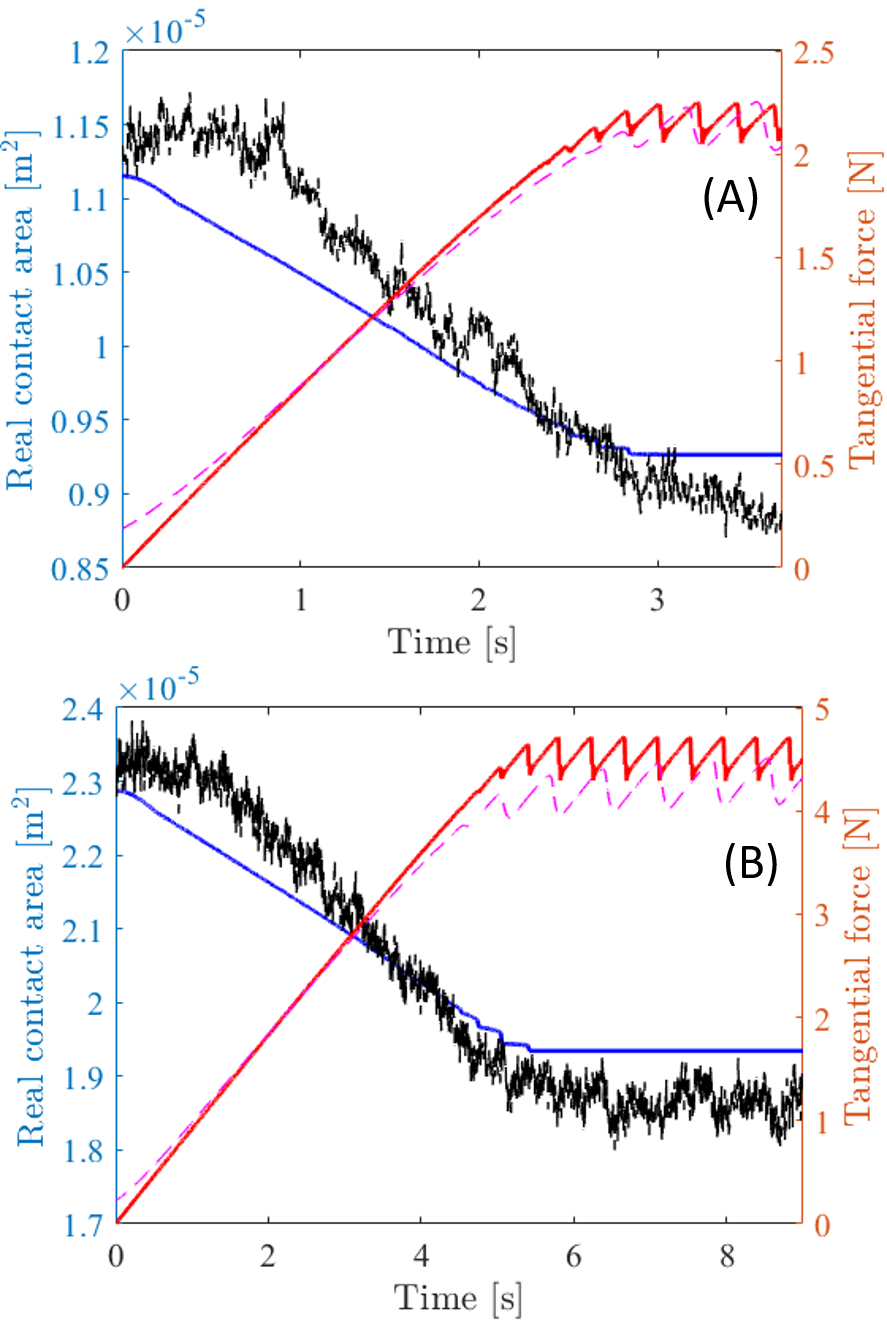}
\caption{Direct comparison between measurements and model predictions, for two typical experiments with either $P$=1.53N (A) or $P$=3.10N (B). Time evolutions of the measured (dashed, black) and predicted (solid, blue) area of real contact, and of the measured (dashed, magenta) and predicted (solid, red) tangential force.}
\label{Fig3}
\end{centering}
\end{figure}

We run the model of section~\ref{sec:model} with the parameter values described in section~\ref{sec:expparam}, for the 6 different PDMS/glass multicontact experiments reported in~\cite{sahli_evolution_2018}. Figure~\ref{Fig3} compares, \rev{for two different normal loads,} the measured time evolution of the area of real contact and tangential force to their corresponding model predictions. \rev{Note that the real contact area is essentially proportional to the normal load, as widely discussed in the contact mechanics modeling literature (see e.g. the reviews~\citep{persson_nature_2005, vakis_modeling_2018}) and confirmed in the experiments discussed here (see Fig.S2 of~\citep{sahli_evolution_2018})}. To facilitate comparison \rev{between model prediction and measurement}, the time origin of the experimental data has been offset by the amount necessary to superimpose the measured and predicted force curves in the central portion of their initial increase. Note that the initial non-linear increase of the measured force is due to the non-vanishing bending stiffness of the steel wire used to pull the slider, when it first bends around a pulley before a significant tension arises along the wire. The apparent difference between the measured and predicted values of the initial area of real contact is due to the above mentioned time offset: the initial predicted value exactly corresponds to the measured value from the first image, but the latter image now corresponds to a negative time and is thus not shown in the figure. The observed difference is of the order of the area measurement noise, presumably due to temporal fluctuations in the illumination and noise in the camera's sensor.

\begin{figure}[tb!]
\begin{centering}
\includegraphics[width=0.7\columnwidth]{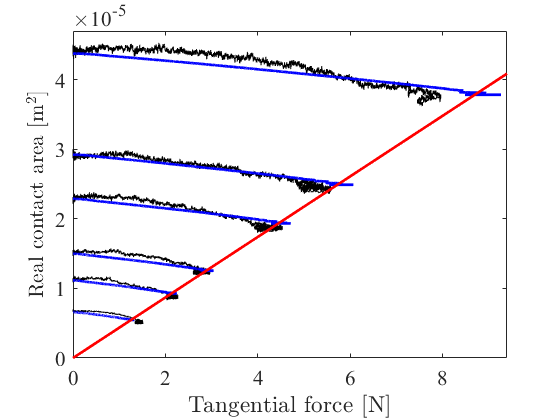}
\caption{Real area of contact \textit{vs} tangential force for the 6 experiments. Black (blue) curves show the measured (predicted) data. The red line has slope $\frac{1}{\sigma_{exp}}$ and passes through the origin.}
\label{Fig4}
\end{centering}
\end{figure}

Figure~\ref{Fig4} then shows the evolution of the area of real contact as a function of the tangential force, for both the measured and predicted data. This figure is similar to Fig.~2A in~\cite{sahli_evolution_2018}, but shows all measurements points rather than just 1 of 130. Note that the stick-slip is responsible for the accumulation of nearly horizontal cycles close to the minimum area/maximum force point of each curve. Also note that the model forces can transiently exceed the value $\sigma_{exp}A$, but always remain smaller than $\sigma A=\frac{\sigma_{exp}}{\epsilon}A$, as expected.

\section{Discussion}\label{sec:discussion}
Although other combinations of model ingredients may have been proposed, we believe that our model incorporates all of the currently available knowledge on the system that we tried to reproduce. As such, it can be seen as the most comprehensive independent asperities model of shear multicontacts so far, to be used for deterministic comparison with the experiments of~\cite{sahli_evolution_2018,sahli_shear-induced_2019}. 

Most of the model parameters ($M$, $k_L$, $v$, $E$, $\nu$, $A_{0i}$, $N$, $\sigma_{exp}$) take their value directly from the measurements. Three adjustable parameters have been systematically varied to choose the most relevant value: $c_{min}$ has no effect on the results, while $\eta$ and $\epsilon$ have been adjusted to reproduce at best the stick-slip regime. Ideally, $p$ and $\alpha_b$ should not be adjustable, but the dispersion in the experimental estimates of $\alpha_i$ is such that their values were not sufficiently constrained. In practice, the value of $p$ was chosen equal to the one suggested from experiments incorporating not only microjunctions within multicontacts, but also millimetric smooth sphere/plane individual contacts~\citep{sahli_evolution_2018}. The value of $\alpha_b$ was then adjusted to best match the overall decay of real contact area during the incipient loading of the interface.

With those values, the time evolution of the tangential load $Q$ is quite well reproduced (see Fig.~\ref{Fig3}). In particular, the slope of the incipient loading is correct, which suggests that the stiffness used for the individual microjunctions is also correct. In contrast, the time evolution of the real contact area is not satisfactory. Of course, the total \textit{amplitude} of the real area decay, from the initial contact to macroscopic sliding, is correct, because we start from the measured initial value ($\sum^N_{i=1} A_{i0}$), and we adjusted $\alpha_b$ to get the correct final value. So we argue that the quality of the comparison between the model and experimental results can only be assessed through the \textit{shape} of the real area decay. And as can be seen from Fig.~\ref{Fig4}, while the shape of the experimental curves $A(Q)$ is essentially quadratic, that of the model curves is much more linear (except from the very beginning, when all microjunctions are pinned and thus decay quadratically according to Eq.~(\ref{eq:areadecay})). We emphasize that this quasi-linear shape is a very robust feature of our model, because we found that the predictions are essentially unaffected by changes in the model assumptions (elliptic \textit{vs} circular microjunctions, Eq.~(\ref{eq:areadecay}) applied at all times or only before the first depinning event) and in the parameter values (for values of $\eta$ and $\epsilon$ enabling stick-slip or not).

The shape of the curve $A(Q)$ results from a sum of a large number ($N$) of complex individual behaviours (non-linear area decay while pinned, constant area while slipping) with distributed parameters (initial area, stiffness, threshold), and is therefore unlikely amenable to a simple explanation. We can however mention an instructive particular case where all microjunctions would have the same initial area. In those (unrealistic) conditions, all microjunctions would behave identically when submitted to a common displacement $X$and thus depin at the same instant. The total area decay would be the sum of $N$ identical quadratic decays, and thus be itself quadratic with the total shear load, until macroscopic sliding. With those specific (but wrong) initial conditions, we would recover a macroscopic area curve with the correct quadratic shape and a simple adjustment of the value of $\alpha_b$ would allow us to provide a good matching with the measurements. This example illustrates the major influence of the distribution of initial areas on the final shape of $A(Q)$. It also clearly shows that the fact that we did not succeed in reproducing a quadratic area decay is not a generic problem of our model, but partly relates to the initial conditions (through the $A_{0i}$) imposed by the experimental dataset.

Could there be other reasons for the failure of our model to reproduce the evolution of the real contact area? The main model ingredient responsible for this evolution is Eq.~(\ref{eq:areadecay}). The first possibility is that, despite the evidence brought in~\cite{sahli_evolution_2018,sahli_shear-induced_2019}, the anisotropic area reduction under shear would not follow a single behaviour law at all junction scales, from millimeter- to micrometer-sized junctions. This possibility is indeed suggested by a recent adhesion-based model of sheared sphere/plane junctions~\citep{papangelo_shear-induced_2019}, where the authors find that the exponent $p$ varies systematically, for a given sphere, with the normal load applied, and hence the initial area. Here we did not try to apply the model of~\cite{papangelo_shear-induced_2019}, because it would require the knowledge of the characteristic radius of curvature of and normal load on each individual microjunction. In contrast, experimental measurements only provide a combination of both quantities, through the area of the microjunction.

We now argue that the solution to the failure of our model will presumably be much more complex than a mere improvement of the form of Eq.~(\ref{eq:areadecay}). The problem may very well be that the predicted individual force $f_i$ is significantly different from the one that really applies on the microjunction. This is substantiated by Fig.~\ref{Fig:microjunctions} which shows the time evolution of the contact area of various microjunctions. Two of them (4 and 5) were selected to show that the time window over which the area decay occurs can be very different: microjunction 4 has not started to shrink yet when the decay of microjunction 5 is already complete. This observation suggests that the individual tangential forces $f_4$ and $f_5$ evolve very differently during the experiment, although they have very similar initial areas and are submitted to the same tangential displacement of the glass substrate. We speculate that such a difference may be the result of elastic interactions between microjunctions, with junctions in a crowed neighbourhood evolving differently from those far from neighbouring junctions\footnote{The slight initial increase in area of junction 4 in Fig.~\ref{Fig:microjunctions} may be due not only to elastic interactions but also to a slight aging due to viscous creep.}.

Those interactions are ignored in our model of independent microjunctions. We thus believe that, in order for an asperity model to have a chance to quantitatively match experiments like those considered here, tangential elastic interactions must be accounted for to describe the shear behaviour of individual microjunctions. Such improved models may incorporate those tangential interactions in ways similar to models already developed for the normal interactions during normal loading of rough surfaces (see \textit{e.g.}~\cite{ciavarella_re-vitalized_2006,afferrante_interacting_2012,braun_propagation_2014,tromborg_slow_2014,braun_seismic_2018}).

\section{Conclusion}\label{sec:conclusion}

We developed an independent asperity model for the incipient loading and onset of sliding of dry multicontact interfaces between a rough elastic solid and a smooth rigid surface. We used it to attempt the first deterministic comparison with experiments which, in addition to the macroscopic loads and displacements, also considers the individual areas of the many microjunctions forming the interface.

The main outcome is that, although we did our best to incorporate experimentally-based behaviour laws, parameter values and initial conditions into the model, it fails to quantitatively reproduce the measurements of~\cite{sahli_evolution_2018,sahli_shear-induced_2019}. Based on observations at the microjunction scale, we suggest that an interesting starting point for future attempts to improve the quantitative deterministic comparison between asperity models and experiments, may be to incorporate a description of the tangential elastic interactions between microjunctions.

Nevertheless, we anticipate that asperity-based friction models, although accounting for tangential elastic interactions, may suffer from the same limitations as their counterparts for purely normal contact (in particular the difficulty to define asperities when a continuum of length scales is involved in the topography, see \textit{e.g.}~\cite{muser_meeting_2017,vakis_modeling_2018}), and may still be unsuccessful to quantitatively match friction experiments. We thus urge for the concurrent development of continuum models suitable to reproduce friction experiments like those of~\cite{sahli_evolution_2018,sahli_shear-induced_2019}.

\appendix
\section*{Appendix 1: Integration of the equation of motion of the slider}\label{app:algo}

\noindent
\textit{Integration scheme for the differential equation giving $X(t)$}

Equation~(\ref{eq:motion}) is integrated by a second order Leapfrog algorithm.
\begin{equation}
\label{eq:gamma}
\gamma(t) = \ddot X(t) = \frac{1}{M}
\left[k_{L} (v t - X) - \sum_{i=1}^{N} f_i \right] 
- \eta \dot X
\end{equation}
is the acceleration of the slider at the beginning of an integration
step. $\gamma(t=0)$ has to be 
calculated at the start of the simulation. In our case, we start with $X(t=0)
= 0$ and $\dot X(t=0)=0$ so that $\gamma(t=0) = 0$.  

Let $dt$ be the time step. The algorithm first updates $X(t)$ according to:
\begin{equation}
  \label{eq:leap1}
  X(t+dt) = X(t) + dt \; \dot X(t) + \frac{1}{2} dt^2 \; \gamma(t).
\end{equation}

With this updated value of $X$ we calculate the new acceleration 
$\gamma(t + dt)$. In the second stage of the algorithm $\dot X$ is
updated according to
\begin{equation}
  \label{eq:leap2}
   \dot X(t + dt) =  \dot X(t) + \frac{1}{2} dt \; \big( \gamma(t) +
\gamma(t + dt) \big),
\end{equation}
and then everything is ready for the next step starting at $t + dt$.

\smallskip
To properly select $dt$ we calculate the elasticity
constant $k_i$ for each contact at the start of the simulation.
The frequency of oscillation of the slider
of mass $M$ under the total force of the contacts is $\Omega_0 = \sqrt{(\sum_{i=1}^N k_i)/M}$ and the corresponding period is
$T_0 = 2 \pi / \Omega_0$. The calculation uses $dt = T_0 / C_t$ with
$C_t = 10^4$. The exact value
of $dt$ depends on the experiment, but a typical value is $dt = 1\,\mu$s. This
value is sufficiently small with respect to all the frequencies entering in
the dynamics of the slider so that even a second order algorithm gives a very
good numerical accuracy. We have however run some calculations with a
4th order Runge-Kutta method~\cite{carnahan_applied_1969}, which is significantly slower,
but has errors that decay as $dt^4$, to check the accuracy of our
calculations. 

\medskip
\noindent
\textit{Algorithm of the subroutine which calculates $\gamma$}

To compute $\gamma(t + dt)$, $X(t+dt)$ and $x_i$ are known. The main point is
to compute all the forces $f_i$ on the junctions. The state of each junction
is recorded with two flags: $\theta_i$ records its instantaneous state,
$\theta_i = 1$ for a pinned junction, $\theta_i = 0$ for a slipping junction,
and $h_i$ keeps track its history, $h_i = 1$ for a junction which has never been slipping switches to $h_i = 0$ the first time the junction starts to slip, when $f_i \ge f_{si}$.

\smallskip
The program scans all the junctions and performs the following steps:

\begin{itemize}
\item Compute $f_i$ for each junction
  
  \smallskip
  $\star$ \textit{if $h_i = 1$}
  the area of the junction depends on $f_i$ according to
  Eq.~(\ref{eq:areadecay}), which determines $\ell_{\parallel i}$ and then
  $k_i\big[A_i(f_i)\big]$ according to Eq.~(\ref{eq:stiffness}). Thus
  \begin{equation}
  \label{eq:fi}
  f_i = k_i\big[A_i(f_i)\big] \; \big(X(t+dt) - x_i\big)
\end{equation}
gives an equation for $f_i$. It is too complex to be solved analytically. We
solve it by an iterative process, using a dichotomy method starting from the
value of $f_i$ from the previous step. Once $f_i$ is
known we update $A_i(f_i)$, $k_i(A_i)$ and $f_{si} = \sigma A_i$ for further
steps.

  \smallskip
  $\star$ \textit{if $h_i = 0$}
  \begin{itemize}
  \item if $\theta_i = 1$ the junction is pinned but $A_i$ is fixed, as well
    as $k_i(A_i)$, and they are known from previous iterations so that $f_i =
    k_i \big( X(t) - x_i \big)$.
  \item if $\theta_i = 0$ the junction is slipping. $f_i = \epsilon f_{si}$.
  \end{itemize}

\item Check for transitions in the junction state
  \begin{itemize}
  \item if $\theta_i = 1$ (pinned junction) then if $f_i \ge f_{si}$ the
    junction starts to slip so that $\theta_i$ switches to 0, $f_i = \epsilon f_{si}$. $h_i$ switches to 0 if it was still equal to 1.
  \item if $\theta_i = 0$ (slipping junction), if $|\dot{X}(t)| < c_{min}
    \times v$ the junction repins, $\theta_i$ switches to 1 and we set
    $x_i = X - \epsilon f_{si}/k_i$ so that the junction starts in the pinned state with $f_i=\epsilon f_{si}$.
  \end{itemize}

\end{itemize}
Once all junctions have been scanned and all $f_i$ are determined, we can compute
$\gamma$ from Eq.~(\ref{eq:gamma}).


\section*{Conflict of Interest Statement}

The authors declare that the research was conducted in the absence of any commercial or financial relationships that could be construed as a potential conflict of interest.

\section*{Author Contributions}

JS and MP designed and analyzed  the results of the model. MP developed the numerical implementation of the model, and produced the model data shown in Figures 3 and 4. RS performed the experiments and produced the experimental data shown in Figures 1, 2, 3 and 4. JS wrote the manuscript. All authors commented on the manuscript.

\section*{Funding}
This work was supported by LABEX MANUTECH-SISE (ANR-10-LABX-0075) of Universit\'e de Lyon, within the program Investissements d'Avenir (ANR-11-IDEX-0007) operated by the French National Research Agency (ANR). It has been supported by CNRS-Ukraine PICS Grant No. 7422.

\section*{Acknowledgments}
We thank Oleg M. Braun, who initiated the work, but passed away before its completion.


\bibliographystyle{frontiersinSCNS_ENG_HUMS}

\end{document}